# Utilization of a Buffered Dielectric to Achieve High Field-Effect Carrier Mobility in Graphene Transistors

*Damon B. Farmer\*, Hsin-Ying Chiu, Yu-Ming Lin, Keith A. Jenkins, Fengnian Xia, and Phaedon Avouris\**

IBM T.J. Watson Research Center, Yorktown Heights, New York 10598, USA.

\* dfarmer@us.ibm.com, avouris@us.ibm.com

We utilize an organic polymer buffer layer between graphene and conventional gate dielectrics in top-gated graphene transistors. Unlike other insulators, this dielectric stack does not significantly degrade carrier mobility, allowing for high field-effect mobilities to be retained in top-gate operation. This is demonstrated in both two-point and four-point analysis, and in the high-frequency operation of a graphene transistor. Temperature dependence of the carrier mobility suggests that phonons are the dominant scatterers in these devices.

Much of the interest surrounding graphene is due to the high carrier mobility that is exhibited by this material.[1,2,3] High intrinsic mobilities in graphene combined with the ability to modulate the carrier density leads to high field-effect mobilities ($\mu_{FE}$) in graphene-based field-effect transistors (FETs).[1] This makes graphene a material of great promise as the active element in electronic devices, particularly those based on low-noise and high-frequency operation. One problem currently hindering the progress of graphene technology is the carrier scattering that has been shown to greatly limit mobility. Much of this scattering is produced by the detrimental interaction between graphene and the gate dielectric material, a necessary component in top-gated FET device architecture. For instance, we have been able to achieve graphene transistor operating frequencies in excess of 20 GHz, but this was done after top-gate dielectric deposition caused an order of magnitude decrease in carrier mobility.[4] The ability to minimize this degradation is paramount to realizing the full potential of graphene devices, and adequate gate dielectrics are needed to accomplish this. In this letter, we present a dielectric stack that does not significantly degrade $\mu_{FE}$, and demonstrate a high-frequency graphene device that utilizes this material.

The dominant sources of scattering in high quality graphene specimens are phonons and charged impurities.[5,6] Utilization of high-κ dielectric material as the gate insulator is expected to partially screen charged impurities and subsequently enhance carrier mobility.[7] However, it has been found that the use of solid high-κ gate dielectrics does not result in the anticipated mobility increase.[8] This is likely due to charge traps in the dielectric and surface phonon scattering caused by the high-κ material at the dielectric/graphene interface, which counteracts any mobility enhancement gained from charged impurity screening.[9] Unlike phonons associated with bulk graphene[10] or graphene ripples,[11] surface phonons are an extrinsic source of scattering that depends on the thermal excitation of contact surfaces.[5,12] This type of scattering is absent in suspended graphene, where external surfaces are eliminated and high mobilities are achieved,[13] but this configuration suffers from electromechanical limitations and does not lend itself to top-gating. It is therefore essential to engineer passivating dielectrics that not only provide decent capacitive coupling, but also minimize charged impurity and surface phonon scattering processes. We achieve this by combining a high-κ dielectric with a low-κ polymer buffer layer.

The polymer NFC 1400-3CP (JSR Micro, Inc.) is employed as the buffer layer in the following experiments. This commercially available polymer is a derivative of polyhydroxystyrene that is conventionally used as a planarizing underlayer in lithographic processes. It can be diluted in propylene glycol monomethyl ether acetate (PGMEA), and spin-coated onto the graphene surface. The dilution and spin speed are adjusted to control the desired thickness and uniformity of the buffer layer. Methyl and hydroxyl groups contained within the polymer structure serve as ideal reaction sites for atomic layer deposition (ALD) of $HfO_2$, the high-κ component of the dielectric stack. ALD of this material is accomplished using tetrakis(dimethylamido)-hafnium and water at a deposition temperature of 125°C.[14] This low temperature deposition process produces $HfO_2$ films with a dielectric constant of κ = 13. In order to be an adequate gate dielectric, the NFC/$HfO_2$ stack must coat the entire gated area of the graphene channel. Since pristine graphene is inert to ALD reactions,[15] it is essential that the NFC layer be continuous on the graphene surface. We have found that a 24:1 dilution (by volume) of PGMEA:NFC is sufficient for accomplishing this. Spinning this solution at a rate of 4,000 rpm for 60 s results in a buffer layer that is approximately 10 nm thick. This thickness is determined on a silicon surface by ellipsometry, and confirmed to be identical on graphene by analysis that is presented later. After curing the buffer layer at 175°C for 5 min to remove residual solvent, 10 nm of $HfO_2$ is deposited onto the buffer layer to complete the dielectric stack. Capacitance analysis of this stack yields a dielectric constant of κ = 2.4 for the buffer layer, which is a reasonable value for this polymer.[16] Source (S), drain (D), back-gate (BG), and top-gate (TG) electrodes of all devices presented herein are defined by electron beam lithography and consist of 0.5 nm Ti/20 nm Pd/30 nm Au metal stacks that are deposited by electron beam deposition. Also, graphene transistors are patterned using conventional plasma etching techniques to define the geometry of the channel and electrode regions. Unless otherwise noted, all measurements are made in vacuum at 300 K, and all exfoliated graphene flakes are on 300 nm SiO$_2$/Si and were purchased from Graphene Industries Limited.

We first examine the impact that buffered dielectric processing has on the transport properties of back-gated graphene devices. The two-point transfer characteristics (Fig. 1a) and transconductances (Fig. 1a inset) of a graphene flake are compared at different stages of the processing described above. The two-point transconductance is defined as $g_m = dI_D/dV_G$, where $I_D$ is the drain current and $V_G$ is the gate voltage. The Dirac voltage ($V_{Dirac}$) of the device before processing is $V_{Dirac} = -3.5$ V, in close proximity to 0 V, which signifies that the graphene is only mildly doped by the supporting 300 nm SiO$_2$ substrate. The doping level changes during processing from highly p-doped after application of the buffer layer ($V_{Dirac} = 42.5$ V), to moderately p-doped after HfO$_2$ ALD ($V_{Dirac} = 13.25$ V). After ALD, the device is subjected to an O$_2$ plasma treatment that is known to etch graphene. A further trend toward neutral doping is exhibited upon plasma treatment ($V_{Dirac} = 5.75$ V), but no evidence of damage to the graphene in the form of a resistance increase or transconductance decrease is observed. This demonstrates that graphene is totally protected from the plasma by the dielectric stack. Beyond changing the doping level, buffered dielectric processing has a minimal effect on the transfer characteristics. Both the minimum current value at $V_{Dirac}$ and the maximum hole transconductance remain within 15% of their initial values. There is a 40% decrease in the maximum electron transconductance, but this is a feature that is not observed in every device, and as later analysis will indicate, is likely associated with doping-induced conductance asymmetry caused by the electrodes.[17,18] Overall, the results obtained using this buffered dielectric are a dramatic improvement compared to other methods of dielectric coating, where more substantial electronic degradation is the norm.[4,19] For comparison, the effects of two such alternative coating methods, NO$_2$ functionalization[20] and oxidized Al deposition,[21] are shown in Figure 1b.

In order to gain quantitative insight into the effect that the buffered dielectric has on carrier mobility, we adopt a model described in Ref. 21 to analyze the two-point resistive behavior of patterned graphene transistors. Figures 2a and 2b show the back-gated resistance and transconductance profiles of one such device before and after buffered dielectric processing. The maximum transconductances translate into device mobilities ($\mu = g_m L/V_D W C_{BG}$) of 4,400 cm$^2$/Vs before and 3,700 cm$^2$/Vs after processing. The corresponding resistance profiles ($R_T$) can be fitted to the equation,[21]

$$R_T = R_c + \frac{L}{W e \mu_{FE} \sqrt{n_o^2 + n^2}} \quad (1)$$

where n is the modulated carrier concentration,[22] e is the electronic charge, and L and W are the length and width of the graphene channel. This allows for extraction of the intrinsic field-effect mobility ($\mu_{FE}$), contact resistance ($R_c$), and residual carrier concentration ($n_o$). As shown in Figure 2a, the measured resistance before processing is in good agreement with the calculated resistance, and we obtain $\mu_{FE}$ = 8,500 cm$^2$/Vs, $R_c$ = 2.2 k$\Omega$ ($R_c W/2$ = 1.65 k$\Omega$ μm), and $n_o$ = 2.25 x 10$^{11}$ cm$^{-2}$. After processing, $R_c$ remains the same, the extracted hole mobility moderately decreases to $\mu_{FE}$ = 7,300 cm$^2$/Vs, and the residual carrier concentration increases to $n_o$ = 2.77 x 10$^{11}$ cm$^{-2}$. It can be seen in Figure 2b that the hole resistance (V - $V_{Dirac}$ < 0 V) can still be accurately modeled, but the electron resistance (V - $V_{Dirac}$ > 0 V) deviates from the calculation at gate voltages much larger than $V_{Dirac}$. This deviation is consistent with the aforementioned transconductance asymmetry, and considering that transport far from $V_{Dirac}$ is dominated by the contacts,[17,18] is another indication of electrode-induced conductance asymmetry. The increase in $n_o$ with decreasing $\mu_{FE}$ is expected, and is most likely due to increased disorder caused by perturbations to the graphene surface during the coating process.[23] As in Figure 1a, $g_m$ and the corresponding values of $\mu$ and $\mu_{FE}$ after buffered dielectric processing are within 15% of their initial values. While the ability to successfully fit the measured resistances assuming a small decrease in $\mu_{FE}$ and a constant $R_c$ lends physical significance to Equation 1, the accuracy of these extracted values must be determined. This is accomplished with four-point analysis, as described below.

Figures 3a and 3b show four-point measurements of a top-gated graphene transistor, where the top-gate voltage is swept at different back-gate voltages. The four-point configuration eliminates effects of the contact electrodes, and allows for a more direct investigation of the intrinsic electronic properties of the graphene channel. The global back-gate electrostatically dopes the entire graphene flake, resulting in a shift of $V_{Dirac}$ with back-gate voltage (Fig. 3a). The conductance at $V_{Dirac}$ remains constant through these sweeps, as do the electron and hole transconductances, which are more symmetric in these measurements than they were in the two-point analysis presented in Figure 1a (Fig. 3b). This confirms that the asymmetry observed in the two-point measurements is not a symptomatic property of the buffered dielectric, but rather originates from the electrodes as ascribed above. By plotting $V_{Dirac}$ as a function of the top-gate and back-gate voltages,[24] the capacitive coupling ratio between the top-gate and back-gate dielectrics is found to be $C_{TG}/C_{BG} = 16$ (Fig. 3a inset).[25] The established value of $C_{BG} = 1.15 \times 10^{-4}$ F/m$^2$ for 300 nm SiO$_2$ therefore gives $C_{TG} = 1.84 \times 10^{-3}$ F/m$^2$ for the buffered dielectric. With the dielectric constants of HfO$_2$ and the buffer layer known, this value is used to confirm that the thickness of the buffer layer is 10 nm on graphene. The top-gated field-effect mobility of this device is found using the relation $\mu_{FE} = (dG/dV_{TG})(L/WC_{TG})$, where $dG/dV_{TG}$ is the differential change in conductance (G) per differential change in top-gate voltage ($V_{TG}$). At a maximum $dG/dV_{TG}$ of 210 µS/V, which corresponds to a total carrier concentration of 5.5 x 11 cm$^{-2}$, the mobility is $\mu_{FE} = 7,600$ cm$^2$/Vs. This is in good agreement with the back-gated field-effect mobility of this device, which is found by similar analysis to be $\mu_{FE} = 7,400$ cm$^2$/Vs, and is one of the highest room temperature values for top-gated graphene reported thus far. The corresponding two-point top-gate resistance of this device is shown in Figure 3c. As with the back-gated resistance profile in Figure 2b, the measured hole resistance agrees reasonably well with the calculated resistance while there is deviation in the electron resistance. With a maximum $g_m$ of 860 nS, the two-point device mobility is $\mu = 4,200$ cm$^2$/Vs. Since this device has the same electrode configuration and originates from the same graphene flake as the device presented in Figure 2, the values of $R_c$ and $n_o$ are assumed to be the same. With these constraints, the best fit to Equation 1 yields a field-effect mobility of $\mu_{FE} = 7,700$ cm$^2$/Vs, consistent with the four-point result of $\mu_{FE} = 7,600$ cm$^2$/Vs. The close agreement between these two methods of mobility analysis serves to validate Equation 1, and in doing so, reinforces the observation that the buffered dielectric allows for high field-effect mobilities in top-gated graphene devices.

To further examine the interaction between graphene and the buffer layer, we fabricate devices in which graphene is directly exfoliated onto a layer of NFC prior to buffered dielectric processing. This leads to a top-gated device configuration in which surface interactions with $SiO_2$ are eliminated, and in effect, allows for a more direct investigation of the buffer layer. Figure 4a shows the top-gated resistance profile of one such device. With $\mu_{FE}$ = 5,000 cm$^2$/Vs, $R_c$ = 2.8 kΩ ($R_cW/2$ = 1.54 kΩ μm), and $n_o$ = 2.1 x 10$^{11}$ cm$^{-2}$, the resistance is in good agreement with Equation 1. The conductance at $V_{Dirac}$ remains relatively constant between 5 K and 300 K (Fig. 4c inset). This behavior is similar to graphene on $SiO_2$ and suspended graphene before current annealing.[11,13] On the other hand, $g_m$ and $\mu_{FE}$ exhibit a pronounced temperature dependence. As seen in Figure 4b, hole transconductance increases by a factor of three as the temperature decreases from 300 K to 5 K. Furthermore, the linear decrease of $\mu_{FE}$ with increasing temperature (Fig. 4c) differs from the nonlinear behavior that is observed when graphene is on $SiO_2$.[5,6] This indicates that phonon scattering is the dominant scattering process over the entire measured temperature range. In contrast, carrier mobility was found to be insensitive to temperature in top-gated devices that employed alternative dielectrics, where charged impurities were the dominant scatterers.[21]

As mentioned above, there is interest in using graphene transistors for high-frequency applications, but mobility degradation caused by top-gate deposition has hindered progress in this area. Figure 5 shows the high frequency, de-embedded current gain of a top-gated graphene transistor that utilizes the buffered dielectric. These measurements were made and subsequent analysis of the data was performed using procedures outlined in Ref. 4. It can be seen that the current gain ($h_{21}$) decreases with increasing frequency (f) at a rate that is close to the ideal of 1/f. This signifies that the buffered dielectric provides a constant, static gate capacitance within the measured frequency range. The cutoff frequency ($f_T$), defined as the frequency at which the current gain is unity, is 9 GHz. It is noted that this device operates in the linear transport regime, where $f_T$ scales with the gate length as $1/L^2$. The gate length for this device is 1.1 μm. By comparison, a gate length of 225 nm was previously required to obtain the same value of $f_T$.[4] The benefit of buffered dielectric incorporation is evident from this result, and it is anticipated that further scaling and optimization will allow for much higher operating frequencies to be achieved. For example, projection of our results to shorter channel lengths gives a value of $f_T = 1$ THz for $L = 100$ nm.

In summary, we have shown that the addition of a low-κ polymer buffer layer between graphene and conventional gate dielectrics helps to minimize mobility degradation in top-gated graphene FETs. Possible reasons for this include the suppression of extrinsic surface phonons by the buffer layer and reduction of the impurity concentration due to inherent properties of the polymer. The dependence of carrier mobility on temperature suggests that phonon scattering is the dominant scattering process that is introduced by the buffered dielectric. This new coating procedure represents a significant improvement over previous efforts, and will hopefully further the advancement of graphene science and technology.

The authors thank B. Ek and J. Bucchignano for expert technical assistance. This work is supported by DARPA under Contract FA8650-08-C-7838 through the CERA program. The views, opinions, and/or findings contained in this article are those of the authors and should not be interpreted as representing the official views or policies, either expressed or implied, of the Defense Advanced Research Projects Agency or the Department of Defense. Approved for Public Release, Distribution Unlimited.

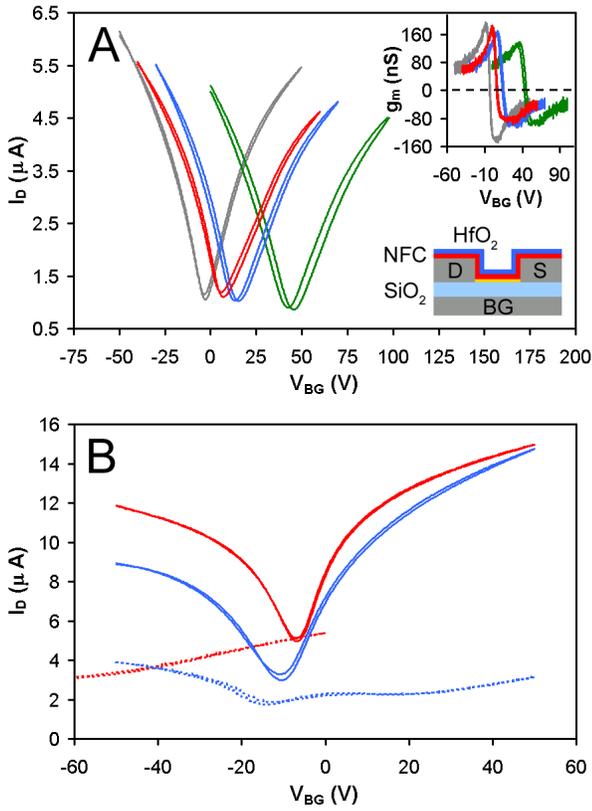

**Figure 1.** Two-point back-gated measurements of graphene flakes. (A) Transfer characteristics and corresponding transconductances (inset) after the different stages of buffered dielectric processing: before processing (grey), after NFC polymer deposition (green), after $HfO_2$ deposition (blue), and after 50 W $O_2$ plasma treatment for 30 s (red). The schematic shows the completed device configuration. (B) Transfer characteristics of two devices before (solid lines) and after (dashed lines) alternative coating processes are employed. 2 nm oxidized Al deposition (red) and $NO_2$ functionalization (blue) are used instead of polymer coating. Both processes result in significant mobility degradation. $V_D = 10$ mV for all measurements, and $V_{BG}$ is swept forwards and backwards to show current hysteresis.

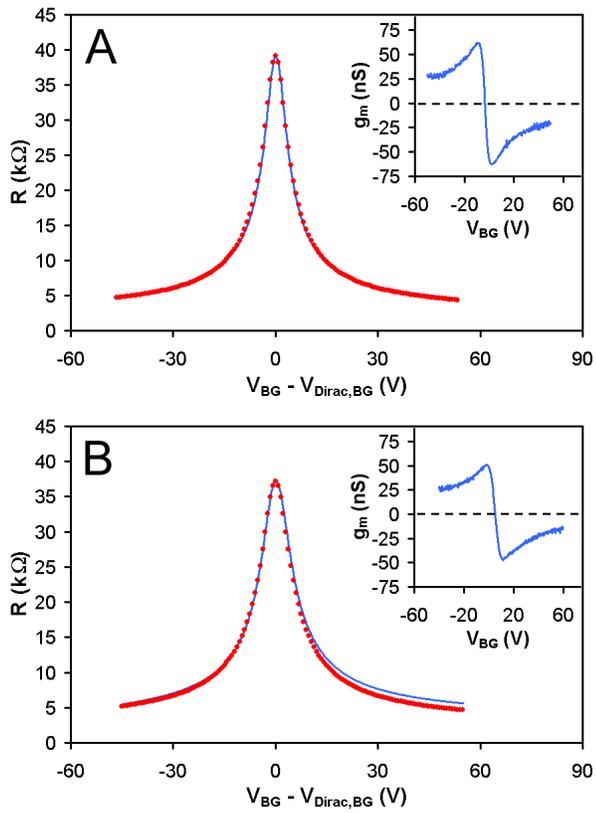

**Figure 2.** Two-point back-gated resistance measurements (blue lines) and calculated results (red circles) with the corresponding transconductances (insets). The field-effect mobility of the device decreases from 8,500 cm$^2$/Vs before (A) to 7,300 cm$^2$/Vs after (B) buffered dielectric processing (L = 17 μm, W = 1.5 μm). $V_D$ = 10 mV for all measurements.

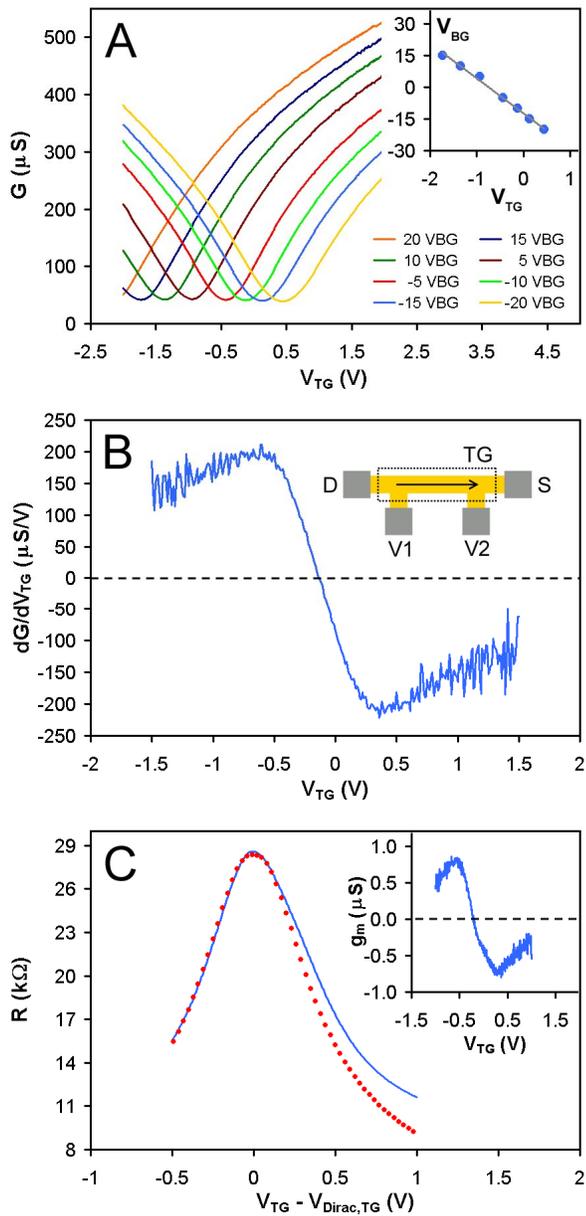

**Figure 3.** Four-point and two-point top-gated measurements of a graphene transistor that incorporates the buffered dielectric. (A) Four-point transfer characteristics at different back-gate voltages. The inset plots $V_{Dirac}$ as a function of $V_{TG}$ and $V_{BG}$. (B) Four-point $dG/dV_{TG}$ for this device. The corresponding maximum field-effect mobility is 7,600 cm$^2$/Vs (L = 10 μm, W = 1.5 μm). Before buffered dielectric processing, this device possessed a back-gated field-effect mobility of 8,300 cm$^2$/Vs. The device configuration is shown in the schematic. For four-point measurements, a constant current of 1 μA is driven between the drain and source while the voltage drop is measured between V1 and V2. For two-

point measurements, the current between the drain and source is measured at a constant drain bias of 10 mV. (C) Two-point measured (blue) and calculated (red circles) resistance of the device, and the corresponding transconductance (inset). The extracted field-effect mobility from this measurement is 7,700 cm$^2$/Vs (L = 13.4 μm, W = 1.5 μm).

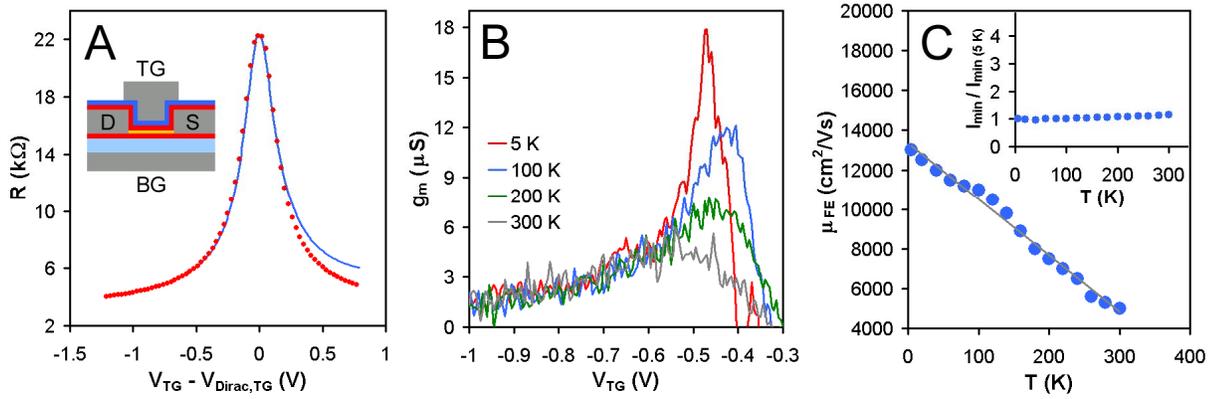

**Figure 4.** (A) Top-gated two-point resistance profile of a device that was exfoliated onto NFC and then capped with the buffered dielectric (L = 0.9 μm, W = 1.1 μm). Fitting the data (blue line) to the resistance equation (red circles) yields a field-effect mobility of 5,000 cm$^2$/Vs at 300 K. The graphene flake that comprises this device was exfoliated from Kish graphite, and the device configuration is shown in the schematic (compare with the schematic in Fig. 1a). (B) Temperature dependence of the transconductance, and (C) corresponding field-effect mobility. The inset shows the normalized current of the Dirac point from 5 K to 300 K. $V_D$ = 10 mV for all measurements.

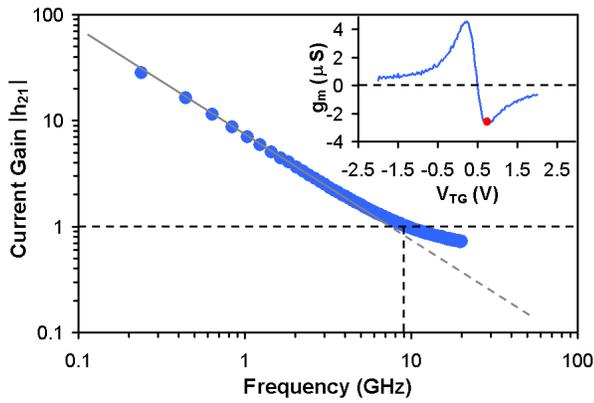

**Figure 5.** Measured current gain ($h_{21}$) of a graphene transistor that incorporates the buffered dielectric as the top-gate dielectric (L = 1.1 μm, W = 16.2 μm). The grey line shows the ideal 1/f dependence, and the intersection of the black dashed lines show $f_T$ = 9 GHz. The inset shows the top-gated transconductance exhibited by this device, and the red circle indicates the point at which the high-frequency measurement was made. $V_{TG}$ = 0.8 V, $V_D$ = 0.5 V, and $I_D$ = 4.59 mA in this measurement.